\documentclass[showpacs,preprintnumbers,amsmath,amssymb]{revtex4}
\usepackage{graphicx}
\usepackage{dcolumn}
\makeatletter
\input{epsf}
\newcommand{\bd}{\begin{document}}
\newcommand{\ed}{\end{document}}
\newcommand{\bc}{\begin{center}}
\newcommand{\ec}{\end{center}}
\newcommand{\bfr}{\begin{flushright}}
\newcommand{\efr}{\end{flushright}}
\newcommand{\vs}{\vspace}
\newcommand{\hs}{\hspace}
\newcommand{\beq}{\begin{equation}}
\newcommand{\eeq}{\end{equation}}
\newcommand{\lb}{\linebreak}
\newcommand{\mb}{\makebox}
\newcommand{\fb}{\framebox}
\newcommand{\mc}{\multicolumn}
\newcommand{\ben}{\begin{enumerate}}
\newcommand{\een}{\end{enumerate}}
\newcommand{\bit}{\begin{itemize}}
\newcommand{\eit}{\end{itemize}}
\newcommand{\un}{\underline}
\newcommand{\lefq}{\lefteqn}
\newcommand{\ba}{\begin{array}}
\newcommand{\ea}{\end{array}}
\newcommand{\beqa}{\begin{eqnarray}}
\newcommand{\eeqa}{\end{eqnarray}}
\newcommand{\beqas}{\begin{eqnarray*}}
\newcommand{\eeqas}{\end{eqnarray*}}
\newcommand{\bfg}{\begin{figure}}
\newcommand{\efg}{\end{figure}}
\newcommand{\bds}{\begin{displaymath}}
\newcommand{\eds}{\end{displaymath}}
\newcommand{\btb}{\begin{tabbing}}
\newcommand{\etb}{\end{tabbing}}
\newcommand{\para}{\parallel}
\newcommand{\pad}{\partial}
\newcommand{\nn}{\nonumber}
\newcommand{\la}{\leftarrow}
\newcommand{\ra}{\rightarrow}
\newcommand{\lgla}{\longleftarrow}
\newcommand{\lgra}{\longrightarrow}
\newcommand{\La}{\Leftarrow}
\newcommand{\Ra}{\Rightarrow}
\newcommand{\Lra}{\Leftrightarrow}
\newcommand{\Lgla}{\Longleftarrow}
\newcommand{\Lgra}{\Longrightarrow}
\newcommand{\bm}{\boldmath}
\newcommand{\lan}{\langle}
\newcommand{\ran}{\rangle}
\renewcommand{\a}{\alpha}
\renewcommand{\b}{\beta}
\newcommand{\g}{\gamma}
\newcommand{\G}{\Gamma}
\renewcommand{\d}{\delta}
\newcommand{\eps}{\epsilon}
\newcommand{\Th}{\Theta}
\newcommand{\s}{\sigma}
\newcommand{\lam}{\lambda}
\newcommand{\D}{\Delta}
\newcommand{\vare}{\varepsilon}
\newcommand{\pr}{\prime}
\newcommand{\ro}{\rho}
\newcommand{\nab}{\nabla}
\newcommand{\m}{\mu}
\newcommand{\n}{\nu}
\newcommand{\Sg}{\Sigma}
\newcommand{\p}{\pi}
\newcommand{\R}{I\!\!R}
\newcommand{\om}{\omega}
\newcommand{\Om}{\Omega}
\newcommand{\ze}{\zeta}
\newcommand{\vart}{\vartheta}
\newcommand{\tri}{\triangle}
\newcommand{\f}{\frac}
\newcommand{\iny}{\infty}
\newcommand{\pro}{\propto}

\begin{document}
\title{Topological Aspects of High Temperature Superconductivity and Berry Phase}

\author{B. Basu}
\email{banasri@isical.ac.in}

\affiliation{Physics and Applied Mathematics Unit\\
 Indian Statistical Institute\\
 Calcutta-700108 }

\begin{abstract}
\centerline{\bf Abstract}
 We propose a mechanism of high
temperature superconductivity from the viewpoint of chirality and
Berry phase. It is observed that spin pairing and charge pairing
is caused by a gauge force generated by magnetic flux quanta
attached to them. From the renormalization group equation
involving the Berry phase factor $\mu$ it is found that there are
two crossovers above the superconducting temperature $T_c$, one
corresponds to the glass phase and the other represents the spin
gap phase. Actually, in this topological framework each charge
carrier is dressed with a magnetic flux quantum and represents a
skyrmion. The skyrmion-skyrmion bound state leads to the d-wave
Cooper pair formation. We have also discussed the Magnus force
acting on the vortices of this system.
\end{abstract}

\maketitle

\newpage

\section{Introduction}
Since high temperature superconductivity \cite{hts} was discovered
in 1986, the investigation of this phenomenon has become one of
the most exciting frontiers of condensed matter theory. A common
feature among the high temperature superconducting compounds is
that the mobility of the charge carriers is almost confined to
planes of Copper-Oxygen atoms, with the off-plane atoms providing
only as a reservoir of carriers. These compounds exhibit
properties of strongly correlated electron systems, being
antiferromagnetic insulators when undoped. It is clear that the
well established BCS theory of low energy superconductivity cannot
provide an adequate description of these new compounds and that
some new approach is needed. Theoretical attempts to explain the
phenomenon have therefore concentrated on strongly correlated
electron models on a 2D lattice.

Immediately after the discovery of high temperature
superconductivity \cite{hts} Anderson \cite{An} proposed a spin
liquid or {\t resonating valence bond}(RVB) state as the theory of
this new phenomenon. Originally the proposal of this RVB state was
for quantum antiferromagnets \cite{An1} on a triangular (or
similarly frustrated ) lattice. Following this proposal Kivelson,
Rokhsar and Sethna \cite{rks} showed that a consequence of the
existence of such a spin
 liquid is that there exist quasiparticles with reversed charge
spin relations; charge 0 spin $1/2$ {\it spinons} and charge $e$
spin 0 {\it holons}. These quasiparticles have topological
character analogous \cite{Lau1,Lau2,bb1} to that of the
quasiparticles in the quantum Hall effect. Recently, there is a
proposal\cite{wen} of algebraic spin liquid (ASL) state with
spin-charge recombination picture to explain the unusual
properties of underdoped high $T_c$ superconductors.

Static or dynamical charge inhomogeneity \cite{zagu,ke,le,ce,ck}
or {\it topological doping} \cite{sak} is a common feature for
doped correlated insulators. In d-dimensions the charge forms
one-dimensional arrays of (d -1) dimensional structures that are
also antiphase domain walls for the background spins. In $d=1$
there is an array of charged solitons \cite{vje} whereas in $d=2$
there are linear {\it rivers of charge} (stripes) threading
through the antiferromagnetic background. In $d=3$ there are
arrays of charged planes \cite{mori}. It is observed in $\m S R$
experiments \cite{wei,nie} that there exists a phase in which
superconductivity coexists with a cluster spin glass. It is
difficult to see how these two phases could coexist unless there
is a glass of metallic stripes dividing the $CuO_2$ planes into
randomly coupled antiferromagnetic regions. In fact, it appears
that the phase diagram of a high-$T_c$ superconductor suggests the
evolution of the system in three stages. Above the superconducting
transition temperature there are two crossovers. The upper
crossover is indicated by the onset of a stripe glass phase. The
lower crossover is where a spin gap or pseudogap (which is
essentially the amplitude of the superconducting order parameter)
is formed. Finally, superconducting phase order is established at
$T_c$.

Emery and Kivelson \cite{ek1} have argued that these experimental
findings support the idea that these self-organized structures are
designed to lower the zero-point kinetic energy. Lee \cite{lee}
observed that the spin-charge separation associated with the
resonating valence bond (RVB) states accounts for all the
qualitative features of the spin gap state. The spins form RVB
singlets so that it costs energy (spin gap) to make triplet
excitations.

Though the spin-charge separation naturally accounts for the
qualitative features of the spin gap state, it has been realized
that there actually exists a strong coupling among spinons and
holons \cite{gb,iof,pal,lbi}  through a gauge interaction and such
a gauge force plays a role essentially to confine spinon and holon
together \cite{frad}. Indeed, spinons and holons are decoupled in
1D and behave just like free particles. However, in 2D the gauge
force plays a crucial role for spin-charge confinement. In the
strong coupling (large U) regime, a correct spin-charge separation
description has been established in a path  integral formalism
\cite{ting} where an electron is described as a composite particle
of a spinon and a holon together with a nonlocal phase-shift
field. It is this phase shift field that helps to recover the
right Fermi surface position.

In this article, we shall study these features from the point of
view of the analysis of high-$T_c$ superconductivity in the
framework of Berry phase \cite{bb2, bb3}. In this scheme the three
dimensional spinons and holons reduce to $\f{1}{2}$ fractional
statistics when the motion is confined to equatorial planes. It is
pointed out that though the spin-charge separation associated with
RVB state can explain well the spin gap state, the superconducting
phase is established when there is spin-charge recombination.
Indeed, the magnetic flux associated with the Berry phase gives
rise to a gauge interaction between spinons and holons which
effectively confines them together. We have investigated the phase
associated with high-$T_c$ superconductivity using renormalization
group fixed point theorem involving the Berry phase factor $\m$
when the Berry phase is given by $e^{i 2 \p \m}$. It is found that
there are three distinct phases: upper phase is associated with
the glass phase at $T_1^*$ and the lower one above the
superconducting transition temperature $T_c$ gives rise
 to the pseudogap (spin gap) at a temperature $T_2^*$. Finally, superconducting
phase is established at $T_c$. It is noted that the spin gap phase
is not independent of the superconducting phase owing to the
manifest presence of the coupling between spin and charge degrees
of freedom. The phase diagrams of different high $T_c$ cuprates
display the universal behavior of   $\f{T_2^*}{T_c}$ as a function
of the hole doping $\f{\delta}{\delta_0}$ with $\delta_0$ being
the optimal doping rate.

In Sec. 2 we shall formulate the basic ideas of the topological
framework of high temperature superconductivity from the viewpoint
of chirality and Berry phase. In Sec. 3 we intend to discuss the
features associated with spin-charge separation and spinon-holon
recombination. In the next section (Sec. 4) with the help of the
remormalization group analysis we shall study the different phases
associated with high temperature superconductivity. In Sec. 5, we
shall show how the skyrmion-skyrmion bound state leads to the
d-wave pairing in this framework. In Sec.6, we shall discuss the
Magnus force acting on the vortices of high temperature
superconductors.

\section{Topological Aspects of Frustrated Antiferromagnets, RVB States
and Berry phase} It is known that in the strong coupling limit and
at half filling the system of correlated electrons on a lattice
which is governed by the Hubbard model can be mapped onto an
antiferromagnetic Heisenberg model with nearest neighbor
interaction and is represented by the Hamiltonian \beq H = J \sum
( S^x_i S^x_j + S^y_i S^y_j + S^z_i S^z_j) \eeq with $J>0$. The
antiferromagnetic model on a triangular lattice emerges as a
frustrated spin system when the  ground state corresponds to the
RVB state. For an antiferromagnetic spin system the existence of
RVB states on a given lattice depends crucially on the type of
lattice which allows frustration to occur. The two characteristic
operators of the ground state of an antiferromagnet,
 namely density of energy
\beq \eps_{ij} = ( \f{1}{4} + \vec{S_i}. \vec{S_j}) \eeq and
chirality \beq W(C) = Tr \prod_{i \in C} ( \f{1}{2} + \vec{\s} .
\vec{S_i}) \eeq ($\s$ are Pauli matrices and $C$ is a lattice
contour) are related \cite{wieg} with the amplitude and phase
$\Delta_{ij}$ of Anderson's RVB through \beq \eps_{ij} = {|
\Delta_{ij} |}^2 \eeq \beq W(C) = \prod_C \Delta_{ij} \eeq This
suggests that $\Delta_{ij}$ is a gauge field. The topological
order parameter $W(C)$ acquires the form of a lattice Wilson loop
\beq W(C) = e^{i \phi (c)} \eeq This is associated with the flux
of the RVB field through \beq e^{i \phi (c)} = \prod_C e^{i
A_{ij}} \eeq where $A_{ij}$ represents a magnetic flux which
penetrates through a surface enclosed by the contour $C$. This is
essentially the Berry phase related to chiral anomaly when we
describe the system in three dimensions through the relation \beq
W(C) = e^{i 2 \pi \m} \eeq where $\m$ appears to be a monopole
strength. In view of this, when a two dimensional frustrated spin
system on a lattice is taken to reside on the surface of a three
dimensional sphere of a large radius in a radial (monopole)
magnetic field we can associate the chirality with the Berry phase
\cite{bb1}. In fact, to take the effect of $spin~ chirality$ in
the RVB theory of high temperature superconductivity we consider a
two dimensional frustrated system in the spherical geometry with a
monopole at the center.

 To study this frustrated spin system leading to
RVB state  characterized by the chirality associated with it we
consider a generalized Heisenberg-Ising Hamiltonian with nearest
neighbor interaction \beq H = J \sum ( S^x_i S^x_j + S^y_i S^y_j +
\Delta S^z_i S^z_j) \eeq where $J > 0$ and the anisotropy
parameter $\Delta = \f{2\m + 1}{2}$ \cite{pb}.  The Berry phase
factor $\m$ can take the values $\m = 0, \pm 1/2, \pm 1, \pm 3/2
........$. It is noted that $\Delta = 1$ corresponds to $\m =
1/2$. Indeed, the Ising part of the Hamiltonian corresponds to the
near neighbor repulsion caused by free fermions and as $\m= 1/2$
is related to a free fermion, we have the isotropic Hamiltonian
which is $SU(2)$ invariant. For $\Delta \rightarrow \infty$, it
corresponds to an Ising system. When $\Delta = 0 (\m = - 1/2)$ we
have the $XX$ model. For a frustrated spin system, this
corresponds to the singlets of spin pairs which eventually
represents the RVB state giving rise to a non-degenerate quantum
liquid. The ground state of antiferromagnetic Heisenberg model on
a triangular lattice which allows frustration to occur is
represented by $\m=-1/2$
 suggesting $\Delta = \f{2\m+1}{2}=0$ in the Hamiltonian (9). Indeed, with
$\Delta =0$, the Hamiltonian effectively corresponds to a bosonic
system represented by singlets of spin pairs which eventually
leads to a resonating valence bond state (RVB).

  To study the spinon and holon
excitations in this frustrated spin system, let us consider a
single spin down electron at a site $j$ surrounded by an otherwise
featureless spin liquid representing a RVB state. As a result, the
state characterized by $|\m|=1$ is formed by the single spin state
$(\m = - \f{1}{2})$ in the spin liquid and the {\it orbital spin}
caused by the {\it monopole} represented by $\mu=-\f{1}{2}$
characteristic of a frustrated spin system leading to RVB ground
state. Thus for the neutral spin $\f{1}{2}$ excitation, the spinon
characterized by $|\m|=1$ may be split into two parts : one spin
$\f{1}{2}$ excitation with $|\m|=\f{1}{2}$ in the bulk and the
other part is due to the {\it orbital spin} by $|\m|=\f{1}{2}$ in
the background characterized by the chirality of a frustrated spin
system. This is analogous to the idea of Laughlin \cite{La} that
spinons obey $\f{1}{2}$ fractional statistics. It may be noted
that such a spinon will be characterized by non-Abelian Berry
phase.

It may be mentioned here that the RVB spin singlet state forming
the quantum liquid are equivalent to FQH liquid with filling
factor $\n = 1/2$ \cite{bb1}. Indeed, in earlier papers
\cite{bb5,bb6} we have pointed out that in QHE the external
magnetic field causes the chiral symmetry breaking of the fermions
(Hall particles) and an anomaly is realized in association with
the quantization of Hall conductivity. This helps us to study the
behavior of a quantum Hall fluid from the view-point of the Berry
phase which is linked with chiral anomaly when we consider a 2D
electron gas of N-particles on the surface of a three dimensional
sphere in a radial (monopole) strong magnetic field. For the FQH
liquid with even denominator filling factor {\it i.e.} for the
state with $\n = 1/2$, the
Dirac quantization condition $e \m = 1/2$ suggests that $\m=1$.\\
In the angular momentum relation for the motion of a charged
particle in the field of a magnetic monopole \beq \bf{J} = \bf{r}
\times \bf{p} - \m \bf{\hat r} \eeq we note that for $\m = 1$ ( or
an integer) we can use a transformation which effectively suggests
that we can have a dynamical relation of the form \beq \bf{J} =
\bf{r} \times \bf{p} - \m \bf{\hat r} = \bf{r^\pr} \times
\bf{p^\pr} \eeq This indicates that the Berry phase which is
associated with $\m$ may be unitarily removed to the dynamical
phase. This implies that the average magnetic field may be taken
to be vanishing in these states. However, the effect of the Berry
phase may be observed when the state is split into a pair of
electrons where each electron in the pair is spin polarized with
the constraint of representing the state $\m = \pm 1/2$. These
pairs will give rise to the $SU(2)$ symmetry as we can consider
the state of these two electrons as a $SU(2)$ doublet. This
doublet of Hall particles for $\n = 1/2$ FQH fluid may be taken to
be equivalent to RVB singlets.

As a hole is introduced into the system by doping, this may
combine with the spinon giving rise to a spinless charged
excitation called holons. Thus holons may also be represented by
$|\m|=1$ which eventually form a pair characterized by a flux
$\phi_0 = hc/2e$. This corroborates with the idea of Laughlin
\cite{Lau1} that a gas of such particles might actually be a
superconductor with charge 2 order parameter. Evidently, just like
spinons, holons will also be characterized by non-Abelian Berry
phase.

\section{Spin and Charge Pairing and Spinon-Holon Interaction}

When a hole is introduced in the concerned system, the spinon with
magnetic flux characterized by $|\m_{eff}|=1$ will interact with
the hole through the propagation of the magnetic flux and
eventually this coupling will lead to
 the creation
of the holon attached with magnetic flux corresponding to
$|\m_{eff}|=1$. Hence, the residual spinon will correspond to
$\m_{eff}=0$. This is realized when the unit of magnetic flux $\m
= - \f{1}{2}$ associated with the single down spin in the RVB
liquid will form a pair with another up spin having $\m = +
\f{1}{2}$ associated with the hole.
   Indeed, a spin pair is formed when the isolated
spin in the RVB liquid will be combined with the spin associated
to the hole. Again, the holon having the effective Berry phase
factor $|\m_{eff}| = 1$ will also eventually form a pair of holes.
As we see from eqn.(11), for any integer $\m$ the Berry phase may
be removed to the dynamical phase and the Berry phase is observed
when the system forms a pair such that the units of magnetic flux
are distributed among the pair. It is noted that the bosonic holon
having $|\m_{eff}| = 1$ and the residual {\it bosonic} spinon
having $|\m_{eff}| = 0$ (which eventually represents a pair)
cannot give the correct statistics for electron when these two
form a composite state. The correct statistics is only achieved
when we introduce a phase associated with a unit of magnetic flux
corresponding to $\m = 1/2$ in this composite system. Thus the
spinon holon recombination along with a phase shift only gives
rise to an electron. This corroborates with the spin-charge
separation description in a path integral formalism \cite{ting}
where an electron is described as a composite particle of a spinon
and holon together with  a nonlocal phase-shift field.

It is now observed that the spin pairing as well as charge pairing
in this scheme occurs through a gauge interaction. In case of spin
pairing, we note that when the units of magnetic flux associated
with the spinon having $|\m_{eff}|=1$ are transferred to the hole,
the residual spinon having $|\m_{eff}| = 0$ eventually forms a
pair of spins having $\m = 1/2$ and $-1/2$. The magnetic flux
associated with each spin will give rise to a gauge force
operating between them. Indeed, we can associate a chiral current
with a spin. When a chiral current interacts with a gauge field,
we have the anomaly which is related to the Berry phase through
the relation \cite{db} \beq \label{chi} q~=2 \m =- \frac{ 1}{2}
\int \pad_\a J^5_\a d^4x =~\f{1}{16 \pi^2}~ Tr~
\int~^*F_{\a\b}F_{\a\b}~d^4 x \eeq where $J^5_\a$ is the axial
vector current $\bar{\psi} \g_\a \g_5 \psi$. $F_{\a\b}$ is the
field strength tensor and $^*F_{\a\b} $ represents the Hodge dual.
 Evidently $q=2\m$ represents the Pontryagin index and
the field $F_{ij} (i,j=1,2,3)$ is associated with the background
magnetic field given by \beq B=-\f{1}{2}\epsilon^{ij}F_{ij} \eeq
 Thus we may
consider that this gauge field is responsible for the spin-pairing
observed in high-$T_c$ superconductivity. The same view will also
be valid for a pair of holes which is eventually formed when the
holon gets its share of magnetic flux having $|\m_{eff}|=1$ from
the spinon. This magnetic interaction is responsible for the hole
pairing which is strong enough to overcome the bare Coulomb
repulsion. This leads to the suggestion that the superconducting
phase order will be established when a spin pair each having unit
magnetic flux and a pair of holes each having unit magnetic flux
interacts with each other through a gauge force. That is, the pair
of holes will be attached to the spin pair such that spin -charge
recombination occurs when each hole
 is attached to a spin site of the spin pair. This ensures that the pseudogap
is roughly of the same size
 as the superconducting gap.

 Mathematically, the spin-charge recombination is
formulated in the spirit of Weng, Sheng and Ting \cite{weng}.
 Indeed, the units of magnetic flux
associated with the Berry phase factor $\m$ may be represented
through a phase \beq
 e^{i 2 \p \m} = \prod_C e^{i A_{ij}}
\eeq where the magnetic flux is associated with the gauge field
$A_{ij}$. We can write the effective Hamiltonian for the system as
$$H_{eff}=H_s+H_h$$ where
\beq H_s = -J_s \sum_{<ij> \s} (e^{i \s A_{ij}}) b^\dagger_{i \s}
b_{j \s} + h.c \eeq with $b_{i \s}(b^{\dag}_{i \s})$ as the spinon
annihilation (creation) operator and $A_{ij}$ represents the
magnetic flux penetrating through a surface enclosed by a contour
$C$ and is given by eqn.(14). Similarly, the  Hamiltonian for the
holon may be written as \beq H_h = -t_h \sum_{<ij>} e^{i( -
\phi^0_{ij} + A_{ij})} h^\dagger_i h_j + h.c \eeq where
$h_i(h^{\dag}_i)$ is the holon annihilation (creation) operators
respectively. Here $\phi^0_{ij}$ represents flux quanta threading
through each plaquette. The interaction between spinons and holons
are then mediated through these gauge fields $A_{ij}$ as
represented in eqn.(14) \cite{bb2,bb3}. It appears that
superconductivity and magnetism are closely related. Indeed,
spinon-holon interaction as well as the pair interaction is found
to be of magnetic origin as the magnetic flux associated with the
Berry phase is responsible for these features.

In fact, the spin gap or pseudogap essentially corresponds to the
superconducting order parameter with the onset of coherence in the
charge degrees of freedom. It is noted that the temperature
$T_2^\ast$ at which the pseudogap is formed is a bit higher than
the superconducting transition temperature $T_c$. The
superconducting state is characterized by spin-charge
recombination which is responsible for the coherent motion of the
pair of holes thus establishing phase coherence.

\section{Renormalization Group Analysis Towards Different Phases
}
We note that the study of different phases associated with
high-$T_C$ superconductivity indicates that above the
superconducting transition temperature there are two crossovers.
The upper crossover is indicated by the establishment of a cluster
spin glass which suggests the existence of a stripe glass phase.
The lower crossover is where a spin gap or pseudogap is formed.
Finally, superconducting phase order is established at $T_c$. We
shall study these crossovers from the view point of
renormalization group (RG) analysis involving the Berry phase
factor $\m$.

Indeed, we know that
 there is a relationship between the
central charge $c$ in conformal field theory and the Berry phase
factor $\m$ \cite{pb}. This relation suggests the generalization
of Zamolodchikov's $c$-theorem \cite{zamo} in $3+1$ dimension
involving $\m$ and formulate $\m$-theorem . It is noted from the
Hamiltonian (9) that the Ising part has the effective coupling
constant $J^{\prime}=J.\f{2 \m +1}{2}$. To study the different
crossovers in the system we may consider the coupling constant as
a function of temperature. We may take $\m$ not to be a fixed
value but dependent on a parameter. Thus, we can consider a
function $\m (\lambda)$ which satisfies\\
1) $\m$ is stationary at fixed points $\m^\ast$ of the RG flow
{\it i.e.}
$\nabla \m (\lambda^\ast) = 0$\\
2) at the fixed points $\m(\lambda^\ast)$ is equal to the Berry
phase factor
$\m^\ast$ of the theory\\
3) $\m$ is decreasing along the infrared (IR) RG flows i,e.
$L\f{\pad \m}{\pad L} \le 0$ where $L$ is a length scale. This
implies that there is a RG trajectory which flows from an
ultraviolet (UV) fixed point $\lambda^\ast_{UV}$ to an IR fixed
point $\lambda^\ast_{IR}$ then one must
have $\m_{UV} > \m_{IR}$.\\

Now let us consider  magnetic flux quanta passing through a domain
$D$ characterizing a length scale $L$ and let a three dimensional
smearing density function $f(a)$ be a positive decreasing function
such that $a \f{\pad f} {\pad a} \le 0$. We now write the
expression for the field strength \beq {[F_{\a\b} (x)]}_D = \int_D
d^3 a f(a) {\tilde F}_{\a\b} (x,a) \eeq

So from the expression which relates the Berry phase factor $\m$
with the chiral anomaly given by \cite{db} \beq 2 \m =
-\f{1}{16\p^2} \int~~^\ast F_{\a\b} (x) F_{\a\b} (x) d^4 x \eeq
 We can write for the {\it flux density}

$$\m = {[\int \tilde{\m} (x) d^4 x]}_D$$
\beq = -\f{1}{32 \p^2} \int d^4 x \int_D d^3 a f(a) ^\ast F_{\a\b}
(x, a) F_{\a\b} (x,a) \eeq ${[\tilde{\m}(x)]}_D$ effectively gives
the smearing of the pole strength over the domain $D$. The
$\m$-function defined above is a pure number but now explicitly
depends on the length scale $L$ characterizing the size of the
domain. Now noting that a global change of scale $L$ for the
off-critical model amounts to a change of the coupling constant
$\lambda^i \to \lambda^i(L)$, the renormalization group flux
equations can be written as \beq L\f{\pad}{\pad L} \lambda^i = -
\b^i \eeq which suggests that \beq -\b^i \f{\pad \m}{\pad
\lambda^i} = L \f{\pad \m}{\pad L} \le 0 \eeq It is noted that
$\m$ takes the usual discrete values of $0, \pm \f{1}{2}, \pm 1,
\pm \f{3}{2} ...$ at fixed points of the RG flows where $\m$ is
stationary and represents the Berry phase factor $\m^\ast$ of the
theory. In terms of energy scale, this suggests that as energy
increases (decreases) $\m$ also increases (decreases). So to study
a critical phenomena, we can associate a critical temperature such
that a standard discrete value of $\m$ corresponding to the Berry
phase factor $\m^\ast$ represents a fixed point of the RG flows.

Now to study the crossovers associated with high-$T_c$
superconductivity, we consider the $3D$ Heisenberg anisotropic
Hamiltonian representing nearest neighbor interaction given by
eqn.(9)
$$ H = J \sum ( S^x_i S^x_j + S^y_i S^y_j + \Delta S^z_i S^z_j)$$
It is noted that the $1D$ relative of this Hamiltonian is given by
\beq
 H = J \sum ( \s_i^x \s^x_{i+1} + \s^y_i \s^y_{i+1} + \f{2 \m+1}{2} \s^z_i
\s^z_{i+1}) \eeq where the anisotropy parameter $\Delta$ is
written in terms of the Berry phase factor $\m$ corresponding to
the $3+1$ dimensional system.

We note that for $0 \le |\Delta| < 1$ we will have three critical
values corresponding to $\m = 0$, $\m = - \f{1}{2}$ and $\m=-1$
 which
represent the fixed points of the RG flows. We associate three
critical temperatures $T^*_1$, $T^*_2$ and $T_c$ with fixed values
of $\m=0$, $\m=1/2$ and $\m=-1$ respectively. However, in a
frustrated spin system, the chirality demands that $\m$ should be
non-zero. So the critical value $\m = 0$ is not achieved and as
such there will be random coupling around the value $\m = 0$. This
will then represent the cluster glass phase at this critical
temperature $T_1^*$.
 Indeed, in this situation, after doping, holes will form a glass of stripes
which may be regarded as a finite piece of electron gas dividing
the $CuO_2$ planes into randomly- coupled antiferromagnetic
regions. When the dopants are immobile, these holes will form
arrays of metallic stripes which are {\it topological}, as they
are antiphase domain walls for the antiferromagnetic background.

The next crossover will be at $ \m=- \f{1}{2}$ corresponding to
the pseudogap (spin gap) phase. Indeed, $\m=-\f{1}{2}$ suggests
that the anisotropy parameter $\D=0$ and the Hamiltonian
corresponds to the $XX$ model which for a frustrated spin system
corresponds to a bosonic system represented by singlets of spin
pairs. This effectively leads to the RVB state and the spin
-charge separation accounts for the qualitative features of the
spin gap state. The pseudogap temperature $T^*_2$ depends on the
doping rate $\d$ and displays nearly a linear decrease with $\d$.

\begin{figure}
\centerline{\epsfxsize=3.5in\epsfbox{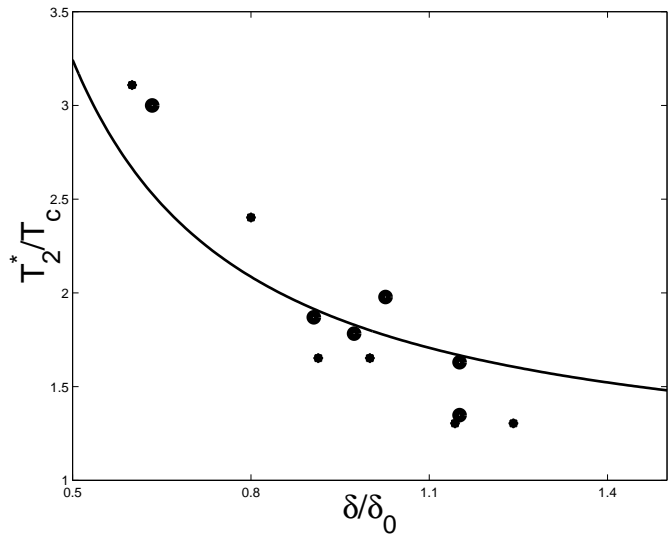}}\caption{The solid
line represents a $T^*_2/T_c$ vs. $\delta/\delta_0$ plot for
$\f{T^*_2}{T_c}=e^{\f{1}{2a}}$ with $a=.85 \times
\f{\delta}{\delta_0}$. Data points for $La_{2-x}Sr_xCuO_4$ and
$Bi_2Sr_2CaCu_2O_{8+\delta}$ are taken from \cite{nakano}.}
\end{figure}

Finally, we have the superconducting transition temperature $T_c$
at $\m=-1$ corresponding to $\D=-\f{1}{2}$. At this point, the
Ising part coupling constant is $\f{-1}{2}J$ with a sign change,
though the bosonic part of the Hamiltonian still dominates with
the coupling constant $J>0$. The sign change of the Ising part is
caused by the presence of the magnetic flux quanta which are
responsible for the interaction between holes in the pair which is
strong enough to overcome the Coulomb repulsion and generates an
attractive force. Indeed, prior to spin-charge recombination, a
spinless holon having $\m=-1$ may be viewed as if a spinless hole
is moving in the background of a monopole characterized by the
strength $\m=-1$. This eventually causes the hole pair formation
each having a magnetic flux quantum characterized by
$\m=\f{-1}{2}$. When the spin-charge recombination occurs a spin
pair each having unit magnetic flux and a pair of holes each
having unit magnetic flux interacts with each other through a
gauge force and a phase coherence is established. It may be noted
that the doping dependence of the spin gap temperature $T^*_2$ and
the superconducting temperature $T_c$ is such that they show
interdependence. The behavior of $\f{T^*_2}{T_c}$ as a function of
$\f{\d}{\d_0}$ where ${\d_0}$ is the optimal doping rate shows a
universal behavior. Indeed, we can derive a relationship between
these two quantities from the following consideration.

From the renormalization group equation (21)
$$ L \f{d \m}{dL} \leq  0,$$ let us specify
\beq L \f{d \m}{dL}=-a \eeq with $a \geq 0$.

Solving this, we find \beq \m = -a~ ln L~+~c \eeq
where $c$ is an arbitrary constant. \\
Now changing the length scale to the temperature $L \sim
\f{1}{T},$ we have \beq \m =a~ln(T)~ +~c \eeq From this for
$\m=-\f{1}{2}$ and $\m=-1$, we get \beq
\f{T^*_2}{T_c}=e^{\f{1}{2a}} \eeq
 Taking $a$ to be a function of
$\f{\d}{\d_0}$, this gives a universal behavior of the dependence
of $\f{T^*_2}{T_c}$ on $\f{\d}{\d_0}$. Indeed taking a simple
ansatz $a=k\f{\d}{\d_0}$with $k$ a constant parameter, we can
compute the respective values. We have found that with $k=0.85$,
the result is consistent with the experimental values obtained for
the high $T_c$ cuprates $Bi_2Sr_2CaCu_2O_8$ and
$La_{2-x}Sr_xCuO_4$ \cite{nakano}.

 We have
obtained three crossovers corresponding to critical points related
to the glass phase, spin gap phase and the superconducting phase.
The temperature $T_2^\ast (> T_c)$ at which the spin gap appears
is found to be dependent on the superconducting transition
temperature $T_c$. This tacitly manifests the presence of coupling
between spin and charge degrees of freedom and the superconducting
phase is characterized by spin-charge recombination.

\section{Skyrmions in high $T_c$ superconductivity and berry phase}
In the present framework, superconductivity arises with the charge
spin recombination when a phase coherence is established. Indeed,
prior to spin-charge recombination, a spinless holon may be viewed
as if a spinless hole is moving in the background of a monopole.
This eventually causes the hole pair formation each having a
magnetic flux quantum characterized by $|\mu|=1/2$. When the spin
charge recombination occurs a spin pair each having unit magnetic
flux interact with each other through a gauge force and a phase
coherence is established.

 Now it is noted that  when a spinless
hole is dressed with a magnetic flux quantum given by $|\mu|=1/2$,
this will represent a skyrmion. Indeed, the magnetic flux quantum
has its origin in the background chirality which is associated
with the chiral anomaly and Berry phase. Indeed, from
eqn.(\ref{chi}), we note that the Berry phase factor $\mu$ is
associated with $^*F_{\alpha\beta}F_{\alpha\beta}$ and we can
write
\begin{eqnarray}
\label{pont}
 q~=&&~2\mu\nonumber\\
 =&&-{\frac{1}{16 \pi^2}}~\int Tr ~^{*} {{F}}_{\alpha\beta} {{F}}_{\alpha\beta} d^4 x\nonumber\\
=&&\int d^4 x~ \partial_\alpha \Omega_\alpha
\end{eqnarray}
where
\begin{equation}
\Omega^{\sigma} = -{\frac{1}{16 \pi^2}} \epsilon^{\sigma\nu \alpha
\beta}
 ~Tr(A_{\nu} F_{\alpha \beta} + {\frac{2}{3}} A_{\nu}A_{\alpha}A_{\beta})
\end{equation}
is the Chern-Simons secondary characteristic class. In case we
have $F_{\alpha\beta}=0$ we can write
\begin{equation}
A_\sigma=g^{-1} \partial_\sigma g,~~~~~~~~~~~     g \in SU(2)
\end{equation}

$\Omega_\sigma$ will represent a topological current $J_\sigma$
given by
\begin{equation}
J_\sigma~=~ {\frac{1}{24 \pi^2}}~ \epsilon^{\sigma\nu\alpha\beta}
~Tr(g^{-1}\partial_{\nu}g)(g^{-1}\partial_{\alpha}g)
(g^{-1}\partial_{\beta}g)
\end{equation}
This may eventually be written in terms of chiral fields
$\pi_a~(a=0,1,2,3)$.
\begin{equation} J_\sigma~=~{\frac{1}{12 \pi^2}}
\epsilon^{\sigma\nu\alpha\beta} \epsilon^{abcd} \pi_a
\partial_{\nu} \pi_b \partial_{\alpha} \pi_c \partial_{\beta} \pi_d
\end{equation}

Now representing a hole by a Dirac fermion field $\psi$ we may
consider the doped hole coupling with the magnetic flux associated
with the chirality in terms of the interaction given by the
Lagrangian
\begin{equation}
L ={\bar \psi}(i{\hat D}+ im(\pi_0 +i\gamma_5 {\vec \pi}{\vec
\tau}))\psi
\end{equation}
where ${\hat D}$=$ \gamma_\sigma(\partial_\sigma - iA_\sigma)$
following the constraint $\pi^2_0+{\vec \pi}^2=1$

The Dirac fermion may be viewed as if it has flavor $N$ so that
for polarized and unpolarized state we have $N=1$ and $2$
respectively. Now integrating for fermions, we can write the
action
\begin{eqnarray}
\label{act}
W~=&&~-~ln~\int exp(-L d^4 x) D\psi~D\bar{\psi}\nonumber\\
=&&~-~N~ln~ {\bf{Det}}(i\hat{D}+im g^{\gamma_5})\nonumber\\
=&&~i~N~\int d^4 x A_\sigma J_\sigma ~ + i\pi N H_3\nonumber\\
&+&NM^2 \int d^4 x ~Tr~ (\partial_\sigma g^{-1}\partial_\sigma g)
\end{eqnarray}
Here $g^{\gamma_5}=\frac{1+\gamma_5}{2} g +\frac{1-\gamma_5}{2}
g^{-1}$. $M$ is a coupling constant having dimension of mass.
$H_3$ is a topological invariant of the map of the space-time into
the target space $S^3$. There are only two homotopy  classes
$\pi_4(S^3)=Z_2$, so that $H_3=0$ or $1$. In fact the term $i \pi
H_3$ is the geometric phase and represents the $\theta$-term. Thus
we see that the charge carriers dressed with magnetic flux can be
represented by a nonlinear $\sigma$-model and may be treated as
skyrmions \cite{super}.

This helps us to view the superconducting pair as
 a skyrmion-skyrmion bound state. Indeed, the
skyrmion excitation is created at each position of the carriers
and plays a role of magnetic field for the carriers. Because of
the magnetic field around a carrier, the Lorentz force acts on
another carrier. Due to this Lorentz force an attractive
interaction is induced between carriers and leads to Cooper pair
formation.

It is noted that the mechanism  suggests a d-wave pairing. As
already pointed out by Kotliar and Liu \cite{kotl} that in the RVB
theory  spinons form the d-wave pairing. Now in the
superconducting pair, the spin charge recombination occurring
through spinon-holon interaction along with the phase coherence
suggests the charge carriers also have d-wave pairing. Indeed, the
fact that superconductivity occurs in the vicinity of
antiferromagnetic long range order, the Cooper pair is d-wave.

To study the underdoped region of cuprates in this framework, we
note that spinon-holon interaction through the gauge force
effectively leads to a spin pair characterized by $\mu_{eff}=0$
where the isolated down spin in the background with $\mu=-1/2$
forms the pair with the up spin of the hole with $\mu=+1/2$.
Indeed this may be taken to represent as a spinon-antispinon bound
state. This essentially corresponds to the SF flux phase as
suggested by Rantner and Wen \cite{wen}. Indeed we can visualize a
spin as a massless fermion and this picture of spinon-holon
interaction may  correspond to a massless fermion coupled to
$U(1)$ gauge field along with the holons coupled with the gauge
field. The pair formed by massless fermions (spins) dressed with
magnetic flux may be viewed as a spinon-antispinon bound state.
This spinon-antispinon bound state present in the nearly
antiferromagnetic chain will enhance the antiferromagnetic
correlation of the system. The simultaneous presence of spin
singlet state will lead to the pseudogap (spin gap). Thus in the
underdoped region we will have the enhancement of the
antiferromagnetic correlation along with the pseudogap. As
mentioned earlier, as doping increases, the antiferromagnetic long
range order is destroyed.

It is known that skyrmion topological defects which are introduced
by doping are responsible for the destruction of the
antiferromagnetic order parameter and their energy may be used as
an order parameter \cite{marn1,marn2}. Indeed, in two spatial
dimensions the nonlinear sigma field $n^a$ may be expressed in the
$CP^1$ Language in terms of a doublet of complex scalar fields
$z_i,~ i=1,2$ with the component $z^{\dag}_i z_i=1$ as
\begin{equation}
n^a=z^{\dag}_i \sigma^a_{ij} z_j
\end{equation}
where $\sigma^a$ are Pauli matrices. In this language the
continuous field theory corresponding to the Heisenberg
antiferromagnet is described by the Lagrangian density in $2+1$
dimensions
\begin{equation}
\label{lns} L_{ns}=(D_\mu z_i)^{\dag} (D_\mu z_i)
\end{equation}
where $D_\mu=\partial_\mu + iA_\mu$ and ${\cal A}_\mu=iz^{\dag}_i
\partial_\mu z_i$. Evidently this possesses solitonic solutions
called skyrmions and charge is defined as
\begin{equation}
Q=\int d^3 x J^0
\end{equation}
where $J^0$ is the zero-th component of the topological current
$J^\mu=\frac{1}{2\pi}\epsilon^{\mu\alpha\beta}\partial_\alpha
{\cal A}_\beta$. It is noted that $Q$ is nothing but the magnetic
flux of the field ${\cal A}_\mu$ indicating that skyrmions are
vortices and represent defects in the ordered Neel state.

Now the following Lagrangian density may be proposed for
describing the dopants and their interaction with the background
lattice in $2+1$ dimensions with the topological $\theta$-term
\begin{equation}
\label{lag} L_{z,\psi}=(D_\mu z_i)^{\dag} (D^\mu
z_i)+i\bar{\psi_a}\partial_\mu \gamma_\mu \psi_a -m^* v_F
\bar{\psi_a} \psi_a -\bar{\psi_a}\partial^\mu \psi_a {\cal A}_\mu
+ L_H
\end{equation}
where the hole dopants are represented by a two-component Dirac
field $\psi_a,~ m^*$ and $v_F$ are respectively the effective mass
and Fermi velocity of dopants. Here $L_H$ is the Hopf term given
by
\begin{equation}
  L_H=\frac{\theta}{2}\epsilon^{\mu\alpha\beta} {\cal A}_\mu \partial_\alpha {\cal A}_\beta
\end{equation}

It is noted that the dopant dispersion relation is given by
\begin{equation}
  \epsilon (k)=\sqrt{k^2 v_F^2 + (m^* v_F^2)^2}
\end{equation}
which is valid for $YBCO~(YBa_2 Cu_3O_{6+\delta})$ where the Fermi
surface has an almost circular shape which is centered at ${\bf k
}=0$. For $LSCO$ $(La_{2-\delta} Sr_\delta CuO_4)$ the Fermi
surface is different \cite{marn2} which corresponds to a
dispersion relation of the form
\begin{equation}
\epsilon (k)= \sqrt{[(k_x \pm \frac{\pi}{2})^2 + (k_y \pm
\frac{\pi}{2})^2] v_F^2+(m^* v_F^2)^2}
\end{equation}
Now following Marino \cite{marn2} the doping parameter $\delta$ is
introduced by means of a constraint in the fermion integration
measure
\begin{equation}
  D[\bar{\psi_a},\psi_a]= D\bar{\psi_a} D \psi_a ~\delta (\bar{\psi_a}\gamma_\mu \psi_a -\Delta^\mu)
\end{equation}
where $\Delta^\mu=4 \delta \int^{\infty}_{x,L} d \xi^\mu~
\delta^3(z-\xi)$ for a dopant at the position $x$ and varying
along the line $L$. Here the factor $4$ corresponds to the
degeneracy of the representation ($4$-component) for the Fermi
fields. This yields the partition function
\begin{eqnarray}
\label{part}
Z=&&~\int D (\bar{z_0},z,{\cal A},\bar{\psi},\psi)~\delta(\bar{z}z-1)~\delta(\bar{\psi}\gamma_\mu \psi-\Delta^\mu)\nonumber\\
&&~\times exp~\{\int^{\infty}_0 d^3 x [~2 \rho_s (D_\mu z^{\dag}_i
D_\mu z_i)+\bar{\psi}(i \partial_\mu \gamma_\mu -\frac{m^* v
_F}{\hbar}-\gamma^\mu {\cal A}_\mu) \psi + L_H]\}
\end{eqnarray}
where $\rho_s$ is the spin stiffness and $L_H$ is the Hopf term.

Upon integration over the fields $z,~\bar{z},~\bar{\psi},~\psi$
the resulting equation of motion for the zero-th component ${\cal
A}_0$ yields the result
\begin{equation}
\label{thet}
  \theta~\epsilon^{ij} \partial_i A_j=4\delta~\delta^2~({\bf
  z}-{\bf x}(t))
\end{equation}
where ${\bf x}(t)$ is the dopant position at a time $t$. If $B$ is
the {\it magnetic flux} or vorticity of ${\cal A}_\mu$ then this
equation becomes
\begin{equation}
\label{theta} \theta B=4\delta~\delta^2~({\bf z}-{\bf x}(t))
\end{equation}

For the skyrmion, $B=\delta^2~({\bf z}-{\bf x}(t))$ indicates that
the skyrmion topological defect configuration coincides with the
dopant position at any time. We see that at zero doping the Hopf
term vanishes. When we translate this result in the $3+1$
dimensional formalism where the $2$D spin system is considered to
reside on the surface of a $3$D sphere with a monopole at the
centre, we note that in the Lagrangian (\ref{lag}), apart from
$\mu$ being a $4$ dimensional index, we have to replace the Hopf
term by the topological Pontryagin term given by
\begin{equation}
  P=-{\frac{1}{16 \pi^2}}~ ^{*} {\mathcal{F}}_{\alpha\beta}
{\mathcal{F}}_{\alpha\beta}
\end{equation}
where
\begin{equation}
\label{star} ^{*}{\mathcal{F}}_{\alpha\beta}
=\frac{1}{2}~\epsilon^{\alpha\beta\sigma\nu}{\cal F}_{\sigma\nu}
\end{equation}

It is noted that in the partition function (\ref{part}) when $\int
L_H d^3 x$ is replaced by $\int P d^4 x$, the latter integral just
represents the Pontryagin index $q$ related to the monopole
strength $\mu$ through the relation $q=2\mu$ as given by
eqn.(\ref{pont}).

From dimensional hierarchy, the relation between topological terms
suggests that in $3+1$ dimensions, when $L_H$ is replaced by
$L_P$, the coefficient $\theta$ is related to $\mu$.

Indeed replacing $L_H$ by the Chern-Simons Lagrangian
\begin{equation}
L_{cs}=\frac{k}{4\pi}\epsilon^{\sigma\alpha\beta} {\cal A}_\sigma
\partial_\alpha {\cal A}_\beta
\end{equation}
We note that the current is given by
\begin{equation}
\label{cur} J_\sigma =\frac{k}{2 \pi}
\epsilon^{\sigma\alpha\beta}\partial_\alpha {\cal A}_\beta
\end{equation}
 and the
zeroth component corresponds to
\begin{equation}
\label{curr} J_0=k\frac{B}{2 \pi}
\end{equation}
So from the relation(\ref{theta}) and (\ref{curr}) and  we find
\begin{equation}
\pi \theta=\frac{k}{2}=2\delta
\end{equation}

It has been shown in ref.\cite{pb} that the Chern-Simons
coefficient $k$ is related to the monopole strength $\mu$ in $3+1$
dimensions by the relation $k=2\mu$. This implies $\mu=2\delta$.
 As in the previous section we have noted
that each charge carrier in the superconducting pair is associated
with the skyrmion topological defect which is caused by the
magnetic flux quantum having $|\mu|=1/2$, superconductivity occurs
at $T=0$ for the critical doping parameter $\delta_{sc}$ given by
$|\mu|=1/2=2\delta_{sc}$ yielding $\delta_{sc}=.25$ for $YBCO$.
When the doping parameter $\delta$ is connected with the oxygen
stoichiometry parameter $x$ we have the  relation $\delta=x-.18$
so that we have $x_{sc}=.43$ \cite{super}, which is in good
agreement with the experimental value $x_{sc}=.41 \pm .02$. For
$LSCO$, the Fermi surface has four branches and this yields
$\delta_{sc}=x_{sc}=.06$ \cite{super} which is to be compared with
the experimental result $x_{sc}=.02$. It is noted that
$\delta_{sc}$ is a universal constant depending only on the nature
of the Fermi surface.

We have pointed out earlier that in $3+1$ dimensions chiral
anomaly leads to the realization of fermions represented by doped
holes interacting with chiral boson fields $\pi_i$, with the
constraint $\pi^2_0+{\vec \pi}^2=1$. The mapping of the space-time
manifold on the target space leads to the homotopy
$\pi_4(S^3)=Z_2$ which takes the values $0$ or $1$ and leads to
the $\theta$-term representing the geometric phase. The third term
in eqn.(\ref{act}) gives rise to the solitonic solution such that
the charge carrier appears as a skyrmion. However in $3+1$
dimensions, the stability of the soliton is not generated by this
term alone as rescaling of the scale variable $x\rightarrow\lambda
x$ may lead to shrinking it to zero size. However, in the present
framework, the attachment of magnetic field with the charge
carrier will prevent it from shrinking it to zero size.

Indeed this gives rise to a gauge theoretic extension of the
extended body so that the position variable may be written as
\begin{equation}
Q_\sigma ~=~q_\sigma~+~iA_\sigma
\end{equation}
where $q_\sigma$ is the mean position. As $\mu=-1/2$ and $+1/2$
corresponds to vortices in the opposite direction we may consider
$A_\sigma$ as $SU(2)$ gauge field when the field strength is given
by
\begin{equation} F_{\sigma\nu} = \partial_{\sigma} A_{\nu} -
\partial_{\nu} A_{\sigma} + [A_{\sigma}, A_{\nu}] \end{equation}
 When $F_{\sigma\nu}$ is
taken to be vanishing at all points on the boundary $S^3$ of a
certain volume $V^4$ inside which $F_{\sigma\nu} \neq 0$, in the
limiting case towards the boundary, we can take
\begin{equation} A_{\sigma} = g^{-1}
\partial_{\sigma } g, ~~~~~~ g \in SU(2) \end{equation}

This helps us to write the action incorporating the $\theta$ -term
as
\begin{equation}
\label{action}
\begin{array}{lcl}
S &=&\displaystyle{\frac{M^2}{16} \int Tr(\partial_{\mu} g^{-1}
\partial_{\mu} g) d^4 x + \frac{1}{32 \eta^2} \int Tr [\partial_{\mu} g
g^{-1}, \partial_{\nu} g g^{-1}]^2 d^4 x}\\
&&\displaystyle{+ {\frac{i\pi}{24 \pi^2}}\int_{S^3} dS_{\mu}
\epsilon^{\mu\nu\lambda\sigma}
Tr[(g^{-1} \partial_{\nu} g)(g^{-1} \partial_{\lambda} g)(g^{-1} \partial_{\sigma} g)]}\\
\end{array}
\end{equation} where $M$ is a constant having the dimension of mass
and $\eta$ is a dimensionless coupling constant. Here the first
term is related to the gauge noninvariant term $M^2 A_{\mu}
A^{\mu}$, the second term (Skyrme term) is the stability term
which arises from the term $F_{\mu\nu}F^{\mu\nu}$ and the third
term is the $\theta$ -term given by $^{*}F_{\mu\nu}F_{\mu\nu}$
which is related to the chiral anomaly and Berry phase.

Marino and Neto \cite{marn2} have pointed out that at the critical
doping $\delta_{sc}$, the energy of the skyrmion vanishes. When we
compute the energy of the skyrmion from the action (\ref{action}),
we find the expression for the minimum energy \cite{skyr} as

\begin{equation}
E_{min}=\frac{12 \pi^2 M}{\eta}
\end{equation}
and the size for $E_{min}$ as
\begin{equation}
R_0=\frac{1}{2M \eta}
\end{equation}
Taking $M$ and $\eta$ as a function of $\delta$, we note that for
the vanishing energy we have $M(\delta_{sc})=0$ which corresponds
to the fact that the spin stiffness vanishes. From the relation
for $R_0$, it indicates that the skyrmion size is infinite.
However, we can have the vanishing energy for finite nonzero
$M(\delta)$ when $\eta$ is infinite. This suggests that at this
point $R_0=0$. This implies that for finite $M$, the vanishing
energy suggests that the skyrmion shrinks to the zero size. So
apart from energy, we can take the size of the skyrmion also as an
order parameter.

\section{Magnus Force}
In this section we shall study the Magnus force in the vortex
dynamics of high $T_c$ superconductors.

It is known that a vortex line is topologically equivalent to a
magnetic flux. Thus in a cuprate superconductor the charge
carriers having magnetic flux associated with them may be viewed
as quantized vortex lines attached to each of them. These vortex
lines lie along the ${\hat z}$ axis. While studying this vortex
dynamics, we assume $T=0$ and low magnetic field so that
vortex-vortex interaction can be ignored. To move a  vortex with
respect to the superconducting flow requires a transverse lift
force which is known as the Magnus force. The Magnus force acting
on a vortex is proportional to the vector product of the velocity
of the vortex relative to the superconducting system and a vector
directed along the vortex core. Ao, and Thouless \cite{ao}
calculated the Berry phase for the adiabatic motion around a
closed loop at zero temperature and showed the existence of the
Magnus force associated with Berry phase, as a general property of
vortex line in a superconductor.

In some recent papers \cite{bb2,bb3} we have
 shown that
due to certain features in the background lattice how Berry's
topological phase plays an important role in describing high $T_c$
superconductivity. Within this framework, we can also study the
Magnus force required for a vortex to move.

In earlier section we have shown that a gauge field is responsible
for the spin pairing and also for the hole pairing. Due to this
interacting magnetic fluxoid  the hole pair can overcome the bare
Coulomb repulsion in high $T_c$ superconductivity. The
superconducting phase is established when spin charge
recombination comes into play $i.e.$ a spin pair with each having
unit magnetic flux and a pair of holes with each hole having unit
magnetic flux interacts with each other through a gauge force.
This gauge field when coupled with the vortex current will lead to
the transverse force responsible for the motion of the vortices.

In our present formalism, we note that in the hole pair the
associated flux quantum corresponding to $|\m|=1/2$ is derived
from the bulk whereas the other flux quantum with $|\m|=1/2$ is
due to the background related to the chirality of the frustrated
spin system. In our model, we may assume that with the movement of
the hole pair, the associated vortex line corresponding to the
contribution from the bulk moves along with the centre of mass of
the paired charge carriers and the condensate will experience an
interaction with the background magnetic field. To study this
interaction, we have to introduce the $\theta-term$ (last term in
the Lagrangian (\ref{stone})) as this corresponds to the vortex
line representing magnetic flux quantum associated with the
background magnetic field. The Lagrangian density of the model in
spherical geometry, where the 2D surface is residing on the
surface of a 3D sphere of large radius with a monopole at the
centre, may be written as \beq \label{stone}
\begin{array}{lcl}
L~&=&\displaystyle{\frac{1}{2}[{\phi}^* ({\partial}_0-ieA_0) \phi-
\phi({\partial}_0+ieA_0) {\phi}^*] +
\frac{1}{2m}{|({\partial}_a-ieA_a)
\phi|}^2 +\frac{\lambda}{2}{({|\phi|}^2-{\rho} _0)}^2+}\\
&&\displaystyle{~~~~~~~~~~~~~\frac{1}{4}F_{\a \b}F^{\a
\b}+\f{1}{4}^* \tilde{F}_{\a\b} \tilde{F}_{\a\b}}\\
\end{array}
\eeq Here $\rho_0$ corresponds to the stationary configuration
with $|\phi|^2=\rho_0$. The term $F_{\a\b}$ corresponds to the
electromagnetic field strength and $\tilde{F}_{\a\b}$ corresponds
to the background magnetic field. $^* \tilde{F}_{\a\b}$ is the
Hodge dual $$^* \tilde{F}_{\a\b}=\f{1}{2}
\epsilon_{\a\b\lam\s}F_{\lam\s}$$ It is noted that the P and T
violating term $^*\tilde{F}_{\a\b}\tilde{F}_{\a\b}$ takes care of
the chirality of the system. It is a four divergence and hence
does not contribute to the equation of motion but quantum
mechanically it contributes to the action. It is noted that there
is a singularity at the z-axis and hence we can take the two
dimensional formalism. To study the vortex dynamics, being
inspired by Stone \cite{stone}, we set $\phi=fe^{i \theta}$ so
that we may write \beq L=if^2({\partial}_0 \theta-ieA_0) +
\frac{f^2}{2m}{({\partial}_a \theta-ieA_a) }^2
+\frac{\lambda}{2}{(f^2-{\rho}_0)}^2+\frac{1}{4}F_{\a \b}F^{\a \b}
+ \f{k}{4 \pi} \epsilon_{\a\b\lam}B_{\a}{\partial}_{\b}B_{\lam}
\eeq It is observed that the dimensional reduction suggests that
the anomalous term $^* \tilde{F}_{\a\b}\tilde{F}_{\a\b}$ in 3+1
dimensions corresponds to the Chern Simons term
$\epsilon_{\a\b\lam}B_{\a}{\partial}_{\b}B_{\lam}$ in 2+1
dimensions.
 We now
introduce Hubbard-Stratonovich fields $\vec{J}$ with the relation
$J_0=f^2$ to obtain \beq\label{hub} L \rightarrow
L^{\prime}=iJ_{\a}({\partial}_{\a} \theta-ieA_{\a}) +
\f{1}{8mJ_0}{{\partial}_a
(J_0)}^2+\f{m}{2J_0}J_a^2+\frac{\lambda}{2}{(f^2-{\rho}_0)}^2+gauge~
field~ terms \eeq We set the vortex part of the phase
$\theta=\bar{\theta}+\eta$ where
$\bar{\theta}=\arg(\vec{r}-\vec{r}_i(t))$ is the singular part of
the phase due to vortices at $\vec{r_i}$ and $\eta$ is the
non-singular part. Integration over $\eta$ suggests the
conservation equation $\partial_{\a}J_{\a}=0$ indicating $J_{\a}$
as a current. So we can identify \beq J_{\a}=
\epsilon_{\a\b\lam}{\partial}_{\b}B_{\lam}=\f{1}{2}
\epsilon_{\a\b\lam} \tilde{F}_{\b\lam} \eeq such that the first
term in expression (\ref{hub}) corresponds to the interaction with
the background magnetic field. Indeed defining the vortex current
\beq K_{\a}=
\epsilon_{\a\b\lam}{\partial}_{\b}{\partial}_{\lam}\bar{\theta}
\eeq we note that the first term in equation (\ref{hub}) can be
written as
$$iB_{\a}(K_{\a}-e \epsilon_{\a\b\lam}{\partial}_{\b}A_{\lam})$$
This shows that the vortex current is coupled to the background
gauge potential $B_{\a}$. It is noted that $J_0$ has an
equilibrium value $\rho_0$ even when the vortex is at rest. Motion
with respect to this background field gives rise to a Lorentz
force which is here just the Magnus force. So the Magnus force is
generated by the background magnetic field when it interacts with
the vortex current. In other words, the Magnus force is generated
by the background magnetic field associated with the chirality of
the system.

To calculate this Magnus force we may take resort to the Berry
phase approach \cite{ao} . When the vortex moves round a closed
loop, we can express the Berry phase $e^{i \phi}$ with \beq
\phi~=~2 \pi N \m \eeq where $N$ is the total number of flux
quantum enclosed by the loop. In our approach each flux quantum in
the background is associated with a hole pair and so the number of
flux quanta $N$ trapped is identical with the number of hole pairs
enclosed by the loop. Thus we can identify $N$ as the number of
hole pairs and we can express $\phi$ as \beq \phi = 2 \pi \m
\f{n_s}{2} \eeq where $n_s$ is the charged superfluid number
density far from the vortex core. The Magnus force is given by the
vector product of the vorticity and the motion relative to the
superconducting velocity \beq F_m~=~ \pm 2 \pi \f{n_s}{2} \m
\hat{c} \times \vec{V}_{vortex} \eeq Here +(-) corresponds to
vortex parallel (antiparallel) to $\hat{c}$ axis and
 $\vec{V}_{vortex}$ is the velocity of
the vortex with respect to the superconducting velocity. It is to
be noted that the Magnus force explicitly depends on the number of
carriers instead of their mass. This supports the Ao, Thouless
theory of the origin of the Magnus force. As high $T_c$
superconductors are type-II superconductors, in the presence of an
external magnetic field, when some magnetic flux quanta penetrates
the material, the number density $n_s$ should be replaced by $n$,
the total density of the fluid when the radius of the integration
contour is much larger than the London penetration depth. This is
a consequence of the Meissner effect \cite{gewe}.

It is known that the Aharonov- Casher phase is generated when the
flux moves through the mobile fluid charges. In the present
situation, the phase arising out of the flux moving through the
fluid charges will be cancelled by that coming from the flux
motion through the static background ion charges. As the net
charge in the macroscopic region is zero, the two Aharonov- Casher
phases will cancel each other.

Actually, the renewed interest on the problem of Magnus force
generated two conflicting points of view on the theory of
transverse force. Volovik \cite{vo} has shown that the motion of
the vortex with respect to the stationary condensate induces a
spectral flow. A momentum transfer from the vortex system to a
heat bath system is caused by a relaxation of the quasiparticles
of the vortex bound states (i.e., the electronic states inside a
vortex core). Therefore the vortex can apparently be moved without
any external source of transverse momentum. In this spectral flow
theory the coefficient of the transverse force $k$ essentially
depends on the electronic states inside a vortex core in
combination of the relaxation time $\tau$ of the quasiparticles.
On the contrary, Ao and Thouless showed that the transverse force
on a moving vortex is a robust quantity which does not depend on
the details of the vortex bound states inside a vortex core but
only on the superfluid density far from the core. The study of
Magnus force in high temperature superconductivity in Berry phase
approach supports the Ao Thouless theory of robust Magnus force.

\section{Discussion}
We have shown above that some characteristic features of
high-$T_c$ superconductivity can be analyzed in the framework of
Berry phase and the different phase structures associated with it
can be well interpreted in this scheme. Emery and Kivelson
\cite{ek1} have argued that the experimental findings support the
idea that the local electronic structures are designed to lower
the zero-point kinetic energy. In view of this, we note that the
Berry phase analysis implicitly determines this condition of
lowering the zero point kinetic energy.

The spin-charge separation associated with the RVB states accounts
for all the features of spin gap state. However, the
superconducting phase is characterized by spin-charge
recombination. Indeed, there exists a strong coupling among
spinons and holons mediated through a gauge interaction and this
gauge force effectively confines spinons and holons together. We
have noted above that the magnetic flux associated with the Berry
phase provides this gauge force and for this no ad-hoc mechanism
is necessary.

The crossovers observed above the superconducting transition
temperature $T_c$ can be well interpreted when we analyze the
anisotropic Heisenberg Hamiltonian using renormalization group
fixed point theorem involving the Berry phase factor $\m$. In our
present formalism it appears that this is a natural consequence
when we consider the formation of spinons and holons in a RVB
state and their interactions in the framework of Berry phase
analysis. In fact, the observed phase diagram in the plane of
temperature $T$ vs. hole doping rate $\delta$ shows the Bose
condensation (superconducting temperature) curve of an {\it arch}
shape rather than the linear increase which manifests the presence
of an {\it optimal doping} \cite{nakano,yas}. However, the
pseudogap temperature displays nearly a linear decrease with
$\delta$. This is in conformity with the experimental findings
which shows an universal behavior of $T_2^\ast/T_c$ as a function
of hole doping $\delta/\delta_0$ with $\delta_0$, the optimal
doping rate.
 This universal
behavior is also manifested in the relationship between
$T^\ast/T_c^{max}$ where $T_c^{max}$ is the maximum
superconducting transition temperature at optimal doping
\cite{nakano,yas}.
 These observations suggest the presence of a relationship
between the spin gap crossover and superconducting phase. Thus the
spin gap phase and the superconducting phase are {\it not
independent} which also manifests the presence of coupling between
spin and charge degrees of freedom \cite{salk}.

In our formalism,  we can make a remark on the mysterious sign
reversal of the
 Hall resistivity (conductivity) effect in the underdoped region in
cuprate superconductors \cite{naga}.  It is noted that in the
underdoped region there will not be sufficient number of holes to
form superconducting pairs. So, in this case,
 a holon characterized by $|\m|=1$ will not be able to share the magnetic
flux with another hole and form the requisite pair. The integral
value of $\m$ will lead to the removal of the Berry phase to the
dynamical phase as given by eqn.(11). Hence the Magnus force will
be decreased. Besides, this in combination with the magnetic flux
lines induced by the external magnetic field within the
penetration depth may change the orientation of the vortices.
Indeed, the interaction of this single holon with $\m=1$ with a
magnetic flux line having $\m=- 1/2$(due to the external magnetic
field)
 will correspond to $\m=1/2$ and as a result we will get a magnetic flux line
with opposite orientation. This change in orientation of the
magnetic flux line will change the sign of the Hall conductivity.
The change of the electronic state due to doping could be related
to the internal electronic structure inside vortex core so that it
affects the dynamic property of vortices. Actually, some people
\cite{haya} have considered this many body effect between vortices
and got results to support the Ao-Thouless theory. In our field
theoretical analysis through Berry phase we got the same result by
calculating the interaction of the background magnetic field with
the vortex current .

The attachment of the magnetic flux of the charge carrier suggest
that this may be viewed as a skyrmion. The interaction of a
massless fermion representing a neutral spin with a gauge field
along with the interaction of a spinless hole with the gauge field
enhances the antiferromagnetic correlation along with the
pseudogap at the underdoped region. The superconducting pairing
may be viewed as caused by skyrmion-skyrmion bound states. This
effectively leads to topological superconductivity.

Abanov and Wiegman \cite{abaw,abaw1} have pointed out that
topological superconductivity in $3+1$ dimensions and $2+1$
dimensions has its roots in the $1$D Peierls-Fr\"{o}hlich model
which suggests that the $2\pi$ phase solitons of the Fr\"{o}hlich
model \cite{fro} are charged and move freely through the system
making it an ideal conductor. In spatial dimension greater than
one this corresponds to superconductivity when the solitonic
feature of a charge carrier is attributed to the attachment of a
magnetic flux to it. It may be remarked here that in $1+1$
dimensions we will have a nonlinear sigma model with the
Wess-Zumino term when the target space is $S^3$ which is the
$O(4)$ nonlinear sigma model. In the Euclidean framework however,
this geometrically corresponds to the attachment  of a vortex line
to the two dimensional sheet which is topologically equivalent to
the attachment of a magnetic flux \cite{royb}. This suggests that
the topological feature of ideal conductivity visualized by
Fr\"{o}hlich in $1+1$ dimensions and that of superconductivity in
$2+1$ and $3+1$ dimensions have a common origin.

\end{document}